\documentclass[conference,compsoc]{IEEEtran}
\IEEEoverridecommandlockouts
\usepackage{booktabs} % For formal tables
\usepackage{subcaption,graphicx}
\usepackage{times}
\usepackage{amsmath}
\usepackage{amssymb}
\usepackage{url}

\ifCLASSOPTIONcompsoc
  \usepackage[nocompress]{cite}
\else
  \usepackage{cite}
\fi

\usepackage[pagebackref=true,breaklinks=true,letterpaper=true,colorlinks,bookmarks=false]{hyperref}

\begin{document}
\title{Differential Geometric Retrieval of Deep Features}

\author{
\IEEEauthorblockN{Y Qian and E Vazquez}
\IEEEauthorblockA{\textit{Cortexica Vision Systems Limited} \\
 30 Stamford Street SE1 9LQ \\
London, UK \\
yu.qian@cortexica.com }
\and
\IEEEauthorblockN{B Sengupta\thanks{YQ and BS contributed equally to this manuscript}}
\IEEEauthorblockA{\textit{Cortexica Vision Systems Limited} \\
\textit{Imperial College London}\\
London, UK \\
b.sengupta@imperial.ac.uk}
}

%\author{Y Qian and E Vazquez\\
%Cortexica Vision Systems Limited\\
%Capital Tower -- 91 Waterloo Road, London SE1 8RT, UK\\
%% For a paper whose authors are all at the same institution,
%% omit the following lines up until the closing ``}''.
%% Additional authors and addresses can be added with ``\and'',
%% just like the second author.
%% To save space, use either the email address or home page, not both
%\and
%B Sengupta\\
%Dept. of Engineering, University of Cambridge\\
%Trumpington Street, Cambridge CB2 1PZ, UK\\
%{\tt\small bs573@cam.ac.uk}
%}

\maketitle

\begin{abstract} 
Comparing images to recommend items from an image-inventory is a subject of continued interest. Added with the scalability of deep-learning architectures the once `manual' job of hand-crafting features have been largely alleviated, and images can be compared according to features generated from a deep convolutional neural network. In this paper, we compare distance metrics (and divergences) to rank features generated from a neural network, for content-based image retrieval. Specifically, after modelling individual images using approximations of mixture models or sparse covariance estimators, we resort to their information-theoretic and Riemann geometric comparisons. We show that using approximations of mixture models enable us to  compute a distance measure based on the Wasserstein metric that requires less effort than other computationally intensive optimal transport plans; finally, an affine invariant  metric is used to compare the optimal transport metric to its Riemann geometric counterpart -- we conclude that although expensive, retrieval metric based on Wasserstein geometry is more suitable than information theoretic comparison of images. In short, we combine GPU scalability in learning deep feature vectors with statistically efficient metrics that we foresee being utilised in a commercial setting.
\end{abstract}

\section{Introduction}
\label{sec:problem}

A common problem in computer vision lies in finding similarity between $2$ (or $3$)-dimensional images (or tensors). This is attained by measuring distances between the two objects, primarily using normalised co-relation, Euclidean distance, Bhattacharyya distance, Jensen-Shannon divergence,  amongst many others. The distances are measured after the images are encoded in some latent space wherein such a latent structure is learnt using a variety of classifiers -- support vector machines (SVMs), logistic regression, etc. Recently, due to the advantages of scalability, large-scale classifier frameworks based on deep-learning have been used for music recommendation \cite{VandenOord2013}, image recommendation \cite{Sengupta2017} as well as general recommendation architectures. 
\cite{Cheng2016}. 
Most of these frameworks do not take the underlying geometry of the feature space into account while making a recommendation. This becomes increasingly important when the similarity between objects is measured in terms of `perception', a quantity that is oblivious to the commonly used distance metrics. The non-trivial problem lies in (a) collecting perceptual similarity between objects in a database (via psychophysics), and (b) using this similarity to construct a metric for classification and retrieval (metric learning, learning to rank, etc.). To compound the problem further, metric used for comparing images might be very different than those for comparing sounds. Regardless of the modality, a large stream of work in neuroscience hypothesise that perception is based on minimising the prediction error between what is observed and what we predict we will observe \cite{Friston2014}. 

In this paper, we start with the $0^{th}$ order problem, i.e., compare how different distance metrics fair against one another when objects that are to be compared are represented as feature spaces induced by a deep neural network; furthermore, we use approximations of the probability density to compute a metric based on the principle of optimal transport \cite{Villani2009} and Riemann geometry \cite{Amari2007}  that takes into account the geometry of transport between two images, an idea that is inherently essential for solving the perceptual similarity problem.
% \cite{Sengupta2016}.

Technically, the problem lies in searching an image database (ranking) with millions of image features ($A_{n_{i} \times r_{m} \times f_{j}}$) for sets of images (say up to 10-20 images) that have similar properties to a query image ($\hat{a}_{query}$). Here, $A$ is a $n_{i} \times  r_{m} \times f_{j}$ tensor where $n_i$ is the total number of images, $r_{m}$ is the length of each feature vector and $f_{j}$ is the total number of features extracted from one image; $\hat{a}_{query}$ is a $r_{m}  \times f_{j}$ query matrix. Although, one can manually construct feature-vectors based on wavelet decomposition, low-rank approximations, etc., we rely on using a convolution neural network (CNN) to compute the feature signature ($r_{m}  \times f_{j}$) for images in the database, as well as the query. 

One way to operationalise a solution lies in weighting each image in the database using a weight vector, and subsequently, extremise the mutual information (or another comparison metric) between the query and the database with respect to the weights. The result of this optimisation problem leads us to a weight vector that provides a rank for all the images in the database when compared to the query image. This is equivalent to measuring distances where each image lies on a continuous probability manifold. There are two contributions of this paper -- (a) in order to describe each image with its deep feature, we use either a computationally efficient approximation of Gaussian Mixture Models (GMMs) or a sparse covariance estimator based on Given rotations, and (b) we provide comparison between these probability distributions using a variety of information-theoretic and geometric metrics.  This work leads us to a much deeper problem where geometric similarity measures can be possibly combined to approximate the metric governing \textbf{perceptual similarity}.

\section{Methods}

\subsection{Dataset and deep-feature generation}
\label{sec:dataset}

In this paper, Describable Textures Dataset (DTD) \cite{Cimpoi2014} is used to evaluate geometric similarity measures for image retrieval. Images in DTD are collected from wild images (Google and Flickr) and classified based on human visual perception \cite{Tamura1978}, such as directionality (line-like), regularity (polka-dotted and chequered), etc. DTD is therefore selected in this research to evaluate the similarity measurements base on human visual perception. DTD contains 5640 wild texture images with 47 describable attributes drawn from the psychological literature and is publicly available on the web at \url{http://www.robots.ox.ac.uk/vgg/data/dtd/}. 

Textures can be described via orderless pooling of filter bank response \cite{Gong2014}. In Deep CNN, the convolutional layers are akin to non-linear filter banks; these have in fact been proved to be better for texture descriptions \cite{Cimpoi2015}. Here, the deep local features are extracted from last convolutional layer of a pre-trained VGG-M \cite{Chatfield2014}. This is represented by $A = \left ( a_{1},...,a_{i},...,a_{N}: a \in \mathbb{R}^{D} \right )$; the size of last convolutional layer is $H\times W\times D$, where $D$ denotes the dimension vector of filter response at the $i^{th}$ pixel of last convolution layer; $N = H\times W$ is the total number of local features. 

For image level representation, two methods are applied to local features -- one is to generate a Gaussian Mixture Model (GMM) model on local descriptors and the second is to estimate a shrunk yet sparse co-variance matrix from the deep feature representation of individual images. A statistical similarity metric is then applied to rank images.  As a baseline for distance calculations, we compute Euclidean distances (${\left\| x \right\|_2} = \sqrt {x_1^2 +  \ldots  + x_n^2}$) between the query image and the database. Similarly, to establish a baseline for feature extraction, we use the Bag of Words (BoW) composed of scale-invariant feature transform (SIFT) features. For further details on SIFT and BoW, please refer to \cite{Sengupta2017}.

\subsection{Retrieval and Ranking}
\label{sec:methods}

Image retrieval using Euclidean norm with bag-of-words feature encoding has been described elsewhere \cite{Sengupta2017}. In subsection \ref{sec:solution1}-\ref{sec:solution4}, we describe three approaches to rank images in terms of their `statistical similarity' (not perceptual similarity).  For the first, we use an information-theoretic divergence while the second and third distances are based on the cost involved in transporting one image to another, and geodesic distance on a Riemannian manifold, respectively. 

To rank images in the database we use two methods, one is to build a Gaussian Mixture Model (GMM) \cite{Murphy2012}, and the second is to estimate a covariance matrix from deep features. For each image, we model the $r_{m}  \times f_{j}$ feature matrix using a 
GMM. Specifically, for computational and analytical efficiency (baseline measure), we approximate the GMM with a Normal distribution, such that the sufficient statistics read,

\begin{eqnarray}
  \tilde \mu  & = & \sum\limits_a {{\omega _a}} {\mu _a} \nonumber \\ 
  \tilde \Sigma  & = & \sum\limits_a {{\omega _a}} \left( {{\Sigma _a} + ({\mu _a} - \tilde \mu ){{({\mu _a} - \tilde \mu )}^T}} \right) 
  \label{eqn:gmm}
\end{eqnarray}

${\mu _a}$, ${\Sigma _a}$ and ${\omega _a}$ are the mean, co-variance and the mixing weights of each Normal distribution (subscript $a$).

The second approximation to an image relies on estimating the co-variance matrix from the feature matrix generated from a deep convolutional neural network. Although the geometry of the co-variance matrix can be utilised to estimate it using low-rank and sparse penalisation, for the sake of computational efficiency, we use an alternative treatment due to \cite{Cao2009,Cao2011}, i.e., a fast sparse matrix transformation (SMT). Briefly, the SMT imposes sparsity constraint on the manifold of co-variance matrices yet maintains a full-rank representation. This is useful as the computation is $\mathcal{O}(f_{j})$; the SMT  can also be seen as a generalisation of FFT and orthonormal para-unitary wavelet transform. 

We will assume that  each feature vector is \textit{i.i.d} zero mean Normal random vectors, and the sample covariance is simply, $\frac{1}{n}A{A^T}$; it is a unbiased estimate of the true covariance matrix, $R = \mathbb{E}[S] = E\Lambda {E^T}$. Often time $S$ is singular, and shrinkage estimators \cite{James1961} are used to regularise the covariance matrix by shrinking it towards a target structure such as an identity matrix, a diagonal matrix with sample variances, amongst others. Sparsity can also be imposed, as in Graphical Lasso \cite{Friedman2008} by imposing a 1-norm constraint on the precision matrix. The maximum likelihood (ML) estimate of the eigenvectors ($E$) and the eigenvalues ($\Lambda$) give us, 

\begin{eqnarray}
  \hat E & = & \mathop {\arg \min }\limits_{E \in {\Omega _k}} \left\{ {\left| {diag\left( {{E^T}SE} \right)} \right|} \right\} \hfill \nonumber \\
  \hat \Lambda  & = & diag\left( {{{\hat E}^T}S\hat E} \right) \hfill 
\label{eqn:MLempirical}
\end{eqnarray}

The SMT constrains the feasible set of ${\Omega _k}$ to a set of orthonormal transformations that are selected as an SMT of order K. A matrix $E$ is an SMT of order K if it can be factorised to K sparse orthonormal matrices, i.e.,

\begin{eqnarray}
  E & = & \prod\limits_{k = 1}^K {{E_k} = } {E_1}{E_1} \ldots {E_k} \hfill  \nonumber \\
  {E_k} & = & I + \Theta ({i_k},{i_j},{\theta _k}) \hfill 
\label{eqn:givensmain}
\end{eqnarray}

Each sparse matrix $E$ can be constructed as a orthonormal Givens rotation on a pair of co-ordinate indexes $({i_k},{i_j})$ of Givens rotations such that,

\begin{eqnarray}
{\left[ \Theta  \right]_{ij}} = \left\{ \begin{gathered}
  \cos ({\theta _k}) - 1,{\text{ if }}i = j = {i_k}{\text{ or }}i = j = {j_k} \hfill \\
  \sin ({\theta _k}),{\text{ if }}i = {i_k}{\text{ and }}j = {j_k} \hfill \\
   - \sin ({\theta _k}),{\text{ if }}i = {j_k}{\text{ and }}j = {i_k} \hfill \\
  0,{\text{ otherwise}} \hfill \\ 
\end{gathered}  \right.
\label{eqn:givenscoeff}
\end{eqnarray}

Using greedy minimization  \cite{Cao2009,Cao2011} we have,

\begin{eqnarray}
  {{\hat E}_k} & = & \arg \min \left| {diag\left( {E_k^T{S_k}{E_k}} \right)} \right| \hfill  \nonumber \\
  {S_{k + 1}} & = & \hat E_k^T{S_k}{{\hat E}_k} \hfill \nonumber \\
   \hfill \nonumber \\
  \hat E & = & \prod\limits_{k = 1}^K {{{\hat E}_k}}  \hfill  \nonumber\\
  \hat \Lambda  & = & diag\left( {{S_{k + 1}}} \right) \hfill  
\label{eqn:givenMLE}
\end{eqnarray}

As a final step, we  obtain a shrunk co-variance matrix where the shrinkage parameter $\alpha$ is selected using cross-validation,

\begin{eqnarray}
  {\Sigma_{SMT}} & = & \hat E\hat \Lambda {{\hat E}^T} \hfill \nonumber \\
  \Sigma & = & \alpha  \cdot {\Sigma_{SMT}} + (1 - \alpha ) \cdot S \hfill 
\label{eqn:shrinkcov}
\end{eqnarray}

\subsubsection{Ranking by KL-divergence}
\label{sec:solution1}

Since there is no analytical solution for the KL-divergence between two GMMs ($V _{a} \sim \mathcal{N}_a(  \omega_{a}, \mu_{a}, \Sigma _a)$ and ($V _{b} \sim \mathcal{N}_b(  \omega_{b}, \mu_{b}, \Sigma _b)$, we utilize two approximations: in the first we approximate the GMM with Eqn. \ref{eqn:gmm}. The KL-divergence ($D_{KL}\left( {{V_a}\left\| {{V_b}} \right.} \right)$) now reads,

\resizebox{1\columnwidth}{!}{
 \begin{minipage}{1.2\columnwidth}
\begin{eqnarray}
\frac{1}{2}\left[ {\log \frac{{\left| {{\Sigma _b}} \right|}}{{\left| {{\Sigma _a}} \right|}} - N_{d} + tr\left( {\Sigma _b^{ - 1}{\Sigma _a}} \right) 
 +  {{\left( {{\mu _b} - {\mu _a}} \right)}^T}\Sigma _b^{ - 1}\left( {{\mu _b} - {\mu _a}} \right)} \right] 
\end{eqnarray}
  \end{minipage}
}

In our experiments, we compute a symmetric-KL divergence which is simply $D_{KL}^{Normal} = \frac{1}{2}{D_{KL}}\left( {{V_a}\left\| {{V_b}} \right.} \right) + \frac{1}{2}{D_{KL}}\left( {{V_b}\left\| {{V_a}} \right.} \right)$. Sorting the KL-divergence provides us with a similarity rank. 

This is a gross-approximation wherein a more subtle approximation relies in bounding the KL-divergence. Particularly, using results from information theory \cite{Hershey2007,Nielsen2016}, we provide retrieval results using a variational approximation to the KL divergence. Particularly, since the log-function is concave, using  Jensen's inequality we have,

\resizebox{1\columnwidth}{!}{
 \begin{minipage}{1.2\columnwidth}
\begin{eqnarray}
{D_{KL}}\left( {{V_a}\parallel {V_b}} \right) & = & {\mathbb{E}_{{V_a}}}\left[ {{V_a}} \right] - {\mathbb{E}_{{V_a}}}\left[ {{V_b}} \right] \nonumber \\ 
  {\mathbb{E}_{{V_a}}}\left[ {{V_b}} \right] & = & {V_a}\log {V_b} \nonumber \\ 
  & = & \sum\limits_a {{\omega _a}\int {{V_a}\log } } \sum\limits_b {{\phi _{b|a}}\frac{{{\omega _b}{V_b}}}{{{\phi _{b|a}}}}}  \nonumber \\
  & \geqslant & \sum\limits_a {{\omega _a}\int {{V_a}\sum\limits_b {{\phi _{b|a}}\log } } } \frac{{{\omega _b}{V_b}}}{{{\phi _{b|a}}}} \nonumber \\ 
   & = & \sum\limits_a {{\omega _a}\sum\limits_b {{\phi _{b|a}}\left( {\log \left( {\frac{{{\omega _b}}}{{{\phi _{b|a}}}}} \right) + \int {{V_a}\log {V_b}} } \right)} }\nonumber   \\
\label{eqn:klapprox}
\end{eqnarray}
 \end{minipage}
}

Here, $\phi _{b|a}$ is a variational parameter that is positive and sums to one. Maximizing \textit{w.r.t} $\phi _{b|a}$ yields,

\resizebox{1\columnwidth}{!}{
 \begin{minipage}{1.2\columnwidth}
\begin{eqnarray}
{\mathbb{E}_{{V_a}}}\left[ {{V_b}} \right] \geqslant \sum\limits_a {{\omega _a}\log \sum\limits_b {{\omega _a}{e^{ - {D_{KL}}\left( {{V_a}\parallel {V_b}} \right)}}} }  - \sum\limits_a {{\omega _a}\mathcal{H}\left( {{V_a}} \right)}
\label{eqn:lowerbound} 
\end{eqnarray}
 \end{minipage}
}

$\mathcal{H}$ is the entropy functional. Subsequently, the variational bound becomes,

\begin{eqnarray}
D_{KL}^{{\text{variational}}}\left( {{V_a}\parallel {V_b}} \right) = \sum\limits_a {{\omega _a}\log \frac{{\sum\limits_{a'} {{\omega _{a'}}{e^{ - {D_{KL}}\left( {{V_a}\parallel {V_{a'}}} \right)}}} }}{{\sum\limits_b {{\omega _b}{e^{ - {D_{KL}}\left( {{V_a}\parallel {V_b}} \right)}}} }}} 
\end{eqnarray}

We symmetrize the variational KL by using $D_{KL}^{\operatorname{var} } = \frac{1}{2}D_{KL}^{\operatorname{var} iational}\left( {{V_a}\left\| {{V_b}} \right.} \right) + \frac{1}{2}D_{KL}^{\operatorname{var} iational}\left( {{V_b}\left\| {{V_a}} \right.} \right)$. Note that such a divergence is the difference of two variational approximations, not a bound in itself.

\subsubsection{Ranking \textit{via} Kantorovich relaxation}
\label{sec:solution3}

%\subsubsection*{Preliminaries} 

Let $\left( {\Psi ,{\psi _m}} \right)$ and $\left( {\Lambda ,{\lambda _m}} \right)$ denote two Polish probability spaces depicting \textit{image 1} and \textit{image 2}, respectively -- ${\psi _m}$ and ${\lambda _m}$. The trivial coupling between the two exists if $\Psi$ and $\Lambda$  are independent so that the coupling is simply a tensor product ${\psi _m} \otimes {\lambda _m}$. A more useful coupling exists when there is a function $S:\Psi  \to \Lambda $ such that $\lambda  = S(\psi )$.  The transport map $S$  is equivalently the change of variables from  ${\psi _m}$ to ${\lambda _m}$. 

\textbf{\small Definition of a transport map:} Let $S$  be a Borel map: $\Psi  \to \Lambda $, the push forward of   ${\psi _m}$ through $S$  is the Borel measure, denoted  ${S_{\# {\psi _m}}}$ defined on $\Lambda $  by ${S_{\# {\psi _m}}}(\Lambda ) = {\psi _m}({S^{ - 1}}(\Lambda ))$. A Borel map: $\Psi  \to \Lambda $   is said to be a transport map if ${S_{\# {\psi _m}}} = {\lambda _m}$. 

In optimal transport \cite{Villani2009}, there is a cost entailed by transporting one measure into another. The transport map then  relies on finding the infimum of  $ \left( {\int_\Psi  {c\left( {x,S(x)} \right)d{\psi _m}:{S_{\# {\psi _m}}} = {\lambda _m}} } \right)$. Optimal transference plans are important because such couplings are stable to perturbations, they encode geometric information about the underlying cost-function, and they exist in smooth as well as non-smooth settings. Given that the existence of this transport map can not be guaranteed, a Kantorovich relaxation amounts to a convex relaxation of Monge's formulation wherein we seek a coupling $\gamma  \in P\left( {\Psi ,\Lambda } \right)$,

\begin{eqnarray}
{\gamma _0} = \mathop {\arg \min }\limits_{\gamma  \in P\left( {\Psi ,\Lambda } \right)} \int\limits_{\Psi  \times \Lambda } {c\left( {{x^\psi },{x^\lambda }} \right)} d\gamma 
\end{eqnarray}

The joint probability measure with the marginals  ${\psi _m}$ and ${\lambda _m}$ allow us to define a Wasserstein distance of order $p$ between  ${\psi _m}$ and ${\lambda _m}$,

\begin{eqnarray}
\mathcal{W}\left( {{\psi _m},{\lambda _m}} \right) = \inf \left( {{{\left\{ {\mathbb{E} \ D{{\left( {{x^\psi },{x^\lambda }} \right)}^p}} \right\}}^{1/p}}} \right)
\end{eqnarray}

$D$ is a distance with a corresponding cost of $c\left( {{x^\psi },{x^\lambda }} \right) = d{\left( {{x^\psi },{x^\lambda }} \right)^p}$. This Earth-Mover or the Monge-Kantorovich distance provides us with a metric over the space of squared integrable probability measures. For two Normal distribution, the $L_2$-Wasserstein distance \cite{Villani2009,Takatsu2008} reads,

\begin{eqnarray}
{D_\mathcal{W}}=\left\| {{\mu _a} - {\mu _b}} \right\|_2^2 + tr\left( {{\Sigma _a} + {\Sigma _b} - 2{{\left( {\Sigma _a^{1/2}{\Sigma _b}\Sigma _a^{1/2}} \right)}^{1/2}}} \right)
\label{eqn:wasser}
\end{eqnarray}

Once the GMMs have been approximated via Eqn. \ref{eqn:gmm} or Eqn. \ref{eqn:shrinkcov}, it is fairly simple to compute distances using Eqn. \ref{eqn:wasser}.

\subsubsection{Ranking \textit{via} Affine Invariant Riemannian Metric}
\label{sec:solution4}

Let us again consider  two feature matrices (query and database), ${V_a} \sim \mathcal{N}(0,{\Sigma _a})$ and ${V_b} \sim \mathcal{N}(0,{\Sigma _b})$. These positive-definite matrices are elements of $S_{ +  + }^{f \times f}$, a space with a defined Riemannian metric \cite{Foerstner2003,Moakher2005}. Under such a geometry, the distance ${D_R}\left( {{V_a},{V_b}} \right)$ between these two matrices is,

\resizebox{1\columnwidth}{!}{
 \begin{minipage}{1.1\columnwidth}
\begin{eqnarray}
D_{R} ({\Sigma _a},{\Sigma _b}) = {\left\| {\log (\Sigma _a^{ - 1/2}{\Sigma _b}\Sigma _a^{ - 1/2})} \right\|_F} = {\left[ {\sum\limits_{c = 1}^C {{{\log }^2}{\lambda _c}} } \right]^{1/2}}
\label{eqn:riemann}
\end{eqnarray}
 \end{minipage}
}

$C$ is the dimension of the co-variance matrix, ${\lambda _c}$ are the eigenvalues, and $F$ represents the Frobenius norm. A useful property of such a distance is that regardless of how the images are manipulated -- be it re-scaling, normalisation, whitening, filtering, etc. -- the distance between the two sources as captured by Eqn. \ref{eqn:riemann} remains invariant. 

To compute Eqn. \ref{eqn:riemann} one can use Eqn. \ref{eqn:gmm} to approximate both GMMs as  Normal distributions; alternatively, the covariance estimated using Eqn. \ref{eqn:shrinkcov} can be used. 

\section{Experiments}
\label{sec:experiments}

In this section, deep feature geometric retrieval methods are evaluated on the DTD dataset. For each image, a set of deep local features is extracted from last convolutional layer of a pre-trained VGG-M. The dimension of each local feature vector is 512. A GMM with 64 components is subsequently generated from the set of local deep features. Normal approximation by GMM and sparse covariance estimation by SMT are used to represent the feature matrix; information-theoretic (Normal and Variational approximation KL) and geometric (Wasserstein and Riemannian) measures to gauge the similarity of two images. In this experiment, Normal approximation KL, Variational approximation KL and Wasserstein metric is applied on GMM model respectively and represented by GMM-Normal KL, GMM-Variational KL and GMM-Wasserstein. Normal approximation KL, Wasserstein and Riemannian metric are applied on sparse covariance generated by SMT respectively and denoted by SMT-Normal KL, SMT-Wasserstein and SMT-Riemannian. 

\begin{table}[ht]
  \begin{center}
    \begin{tabular}{|c|c|c|c|c|}
        \hline
        MAP                                     & Top-1 & Top-5 & Top-10 &  Time \\ \hline
        GMM-Normal KL & 0.53  & 0.46  & 0.42   & 0.375s     \\ \hline
        GMM-Variational KL & 0.45  & 0.42  & 0.38   & 0.016s          \\ \hline
        GMM-Wasserstein                    & 0.62  & 0.52  & 0.46   & 5.147s     \\ \hline
        SMT-Riemannian                       & 0.50  & 0.44  & 0.39   & 0.754s      \\  \hline
        SMT-Normal KL             & 0.53  & 0.44  & 0.39   & 0.125s
        \\  \hline
        SMT-Wasserstein                       & 0.59  & 0.51  & 0.46   & 9.04s
        \\  \hline
                SIFT-BoW-Euclidean                       & 0.43  & 0.37  & 0.32   & 0.0007s
        \\  \hline
    \end{tabular}
  \end{center}
  \caption{Retrieval results on the DTD dataset. Note that VGG-M has been pre-trained on Imagenet.}
  \label{table:DTD}
\end{table}

\begin{figure}
 \centering \includegraphics[scale=0.7]
 {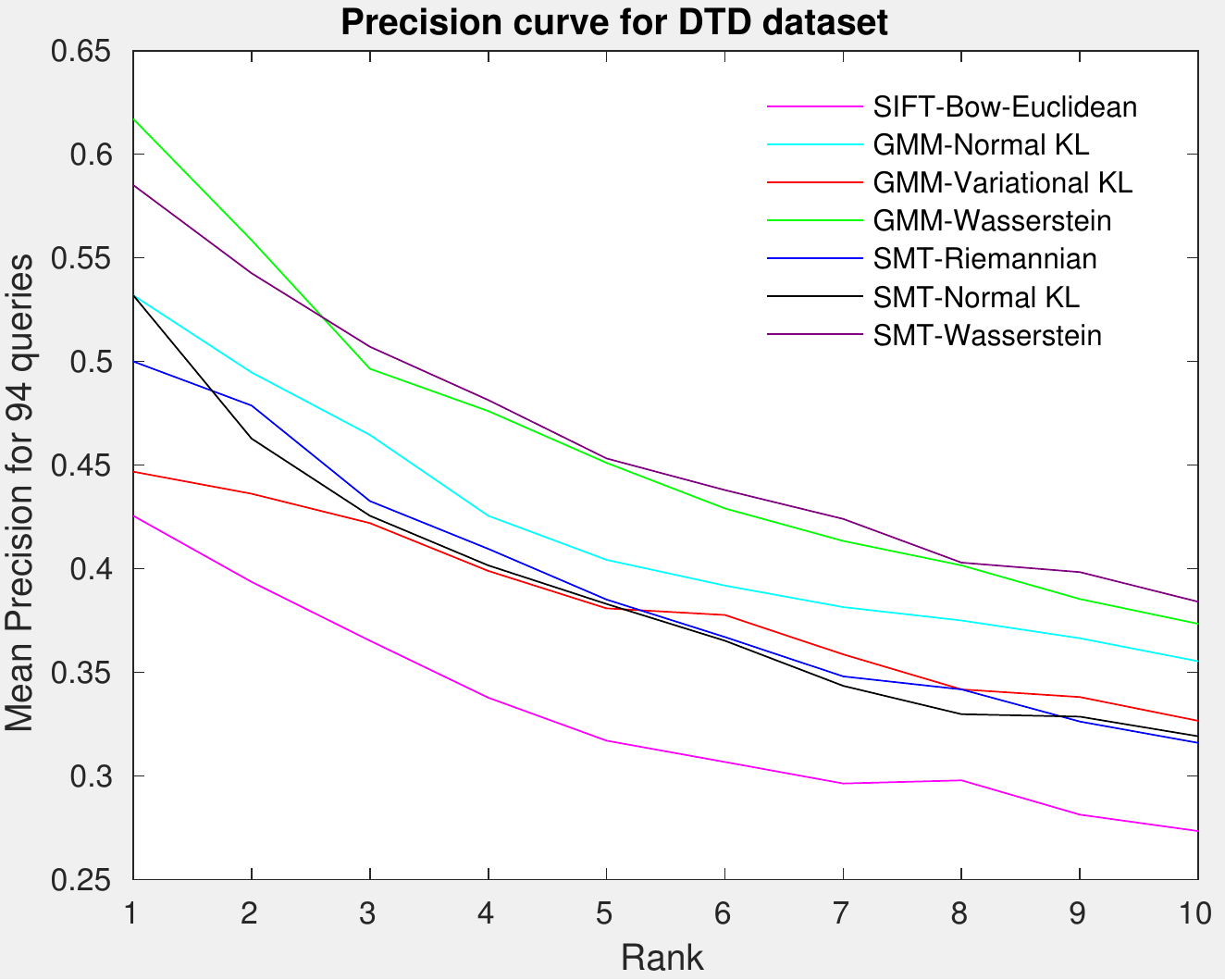}
 \caption{Precision on the DTD dataset}
 \label{fig_DTD_JointAnno}
\end{figure}

\begin{figure}
 \centering \includegraphics[width=0.5\textwidth, height=1.5in]
 {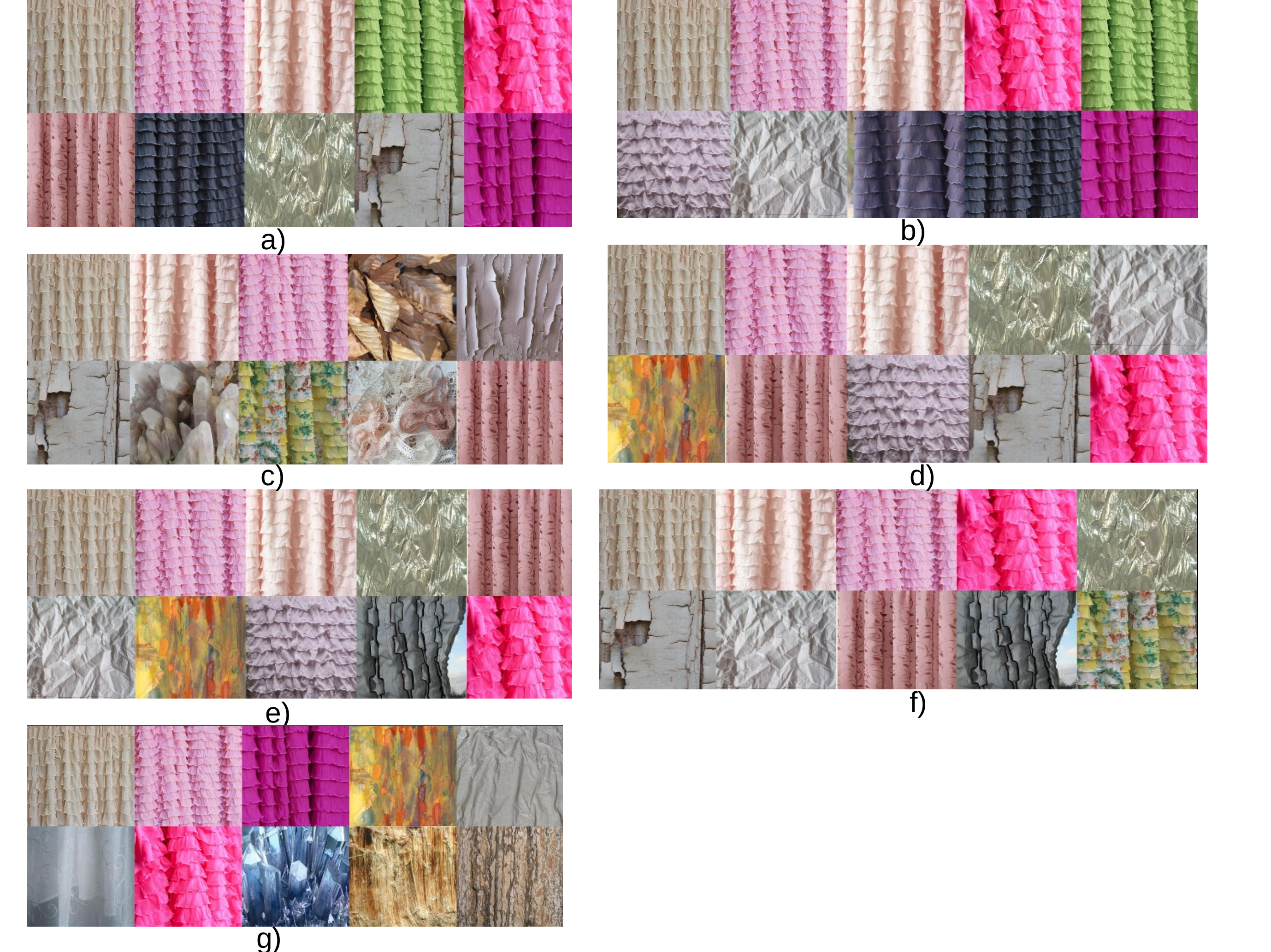}
 \caption{Retrieved results on DTD: a) GMM-Wasserstein  b) SMT-Wasserstein c) GMM-Normal KL d) SMT-Normal KL  e) SMT-Riemannian f) GMM-Variational KL g) SIFT-BoW Euclidean}
 \label{fig_RetrievalDTD}
\end{figure}

To evaluate similarity metric for image retrieval, Mean Average Precision (MAP) on top 10 rankings are calculated. 2 images per category, i.e., a total of 94 images are selected as queries from the test dataset. The dataset retrieved includes 3760 images from DTD training and validation datasets. MAP on DTD is listed in Table \ref{table:DTD}. Average precision on top 10 ranking is displayed in Figure \ref{fig_DTD_JointAnno}. An example of the retrieval obtained with each method is shown in Figure \ref{fig_RetrievalDTD}.  On each case, ten images are displayed. Top left image is the query used. The rest of images are arranged by similarity to query image as obtained with each method.

As is apparent from Table \ref{table:DTD}, computing Wasserstein distances (using GMM or shrank co-variances) prove to be superior when compared to the other methods evaluated, including the baseline that uses a bag of words based SIFT feature generation and Euclidean distances. Surprisingly, variational KL is the least precise distance metric even in contrast to the case where we approximate the feature matrix with a univariate Normal distribution. A possible cause for such a lower precision might be due to the inferior conditioning of the co-variance matrix. 

Query times are shown in Table \ref{table:DTD}; they have been obtained as the average time to calculate the similarity between two images. The code was implemented in Matlab 2015a under Linux with Intel(R) Xeon(R) CPU E5-2640 @ 2.00GHz and 125G RAM.

\section{Conclusion}
\label{sec:concl}

Using hand-crafted features such as scale-invariant feature transform (SIFT), a histogram of oriented gradients (HoG), etc. with Euclidean distances has had a long standing history in computer vision, especially before the advent of deep-learning based feature extractors. Hand-crafted features have poor generalisation capabilities along with being non-robust to non-linear transformations. The same goes for Euclidean distance, which is often not the ambient geometry for the objects being compared. For probability measures, the notion of an ambient geometry is clear due to the Riemann manifold inherited by these measures. In this paper, we have touched upon the $0^{th}$ order problem that may lead to understanding `perceptual similarity'. More specifically, we have used a convolution neural network (CNN) to obtain feature matrices; utilising either Gaussian Mixture Models (GMMs) or shrunk covariance estimators to obtain a probabilistic representation of the features. Subsequently, using information theoretic divergences and Riemann geometric metrics, we compare (dis) similarities between images.

Based on evaluation for DTD dataset, Wasserstein distances show increased retrieval fidelity based on both probabilistic representations, albeit they are more expensive to evaluate.  We believe that the increased accuracy of the Wasserstein distance is due to two properties -- first, the metric does not include calculating the inverse of covariance matrices, thereby enclosing the cases with singularity; in contrast, the KL-divergence between two distributions could easily reach infinity if the covariance of the second distribution becomes singular. The second property, which we hypothesise, is the increased statistical robustness of the Wasserstein distance, i.e., the metric might have small variance when comparing distributions that are closely situated in the parametric manifold.

Although, we have utilised the final convolutional layer of a CNN to distinguish images; much empirical work has shown that there are many general features of an image or a video that are captured by the initial layer of a CNN \cite{Yosinski2014}. By visualising different layers in \cite{Mahendran2015}, it is apparent that the lower layer of CNN can capture more colour information, the higher layers, on the other hand, are more objective. The retrieval result in Figure \ref{fig_RetrievalDTD} demonstrates that colour is not adequately captured due to deep local features extracted from the last convolutional layer, which keeps less colour related information. Hence, the fidelity to distinguish images using any of our retrieval criteria should undoubtedly increase with additional `independent' feature vector that can be computed via the initial or the middle (general to a more specific characterisation of the image) layers of a CNN. Bayesian model averaging or multi-kernel learning, as has been utilised for video-based action recognition might be a way forward \cite{Sengupta2017a,Sengupta2017b}.

\subsection*{\small Factors that affect the successful deployment}
For a commercial system, speed is an essential ingredient. In fact, computing Wasserstein and Riemann distance have their issues. For example, Wasserstein distance in computer vision was proposed more than a decade ago \cite{Rubner1998}. The cost of computing optimal transport between two distributions of dimension $d$ is at least $\mathcal{O}({d^3}\log d)$. This is especially not plausible to compute in a commercial environment when feature vectors are generated by deep convolutional neural networks, which are by construction high dimensional. In our study, even after approximating the GMMs as multivariate Normal distributions, the computational inefficiency is inherent, as computing Eqn. \ref{eqn:wasser} proves to be most expensive amongst all the metrics that we compare.  A solution emerges in the form of low dimensional embedding of the metric space \cite{Grauman2004,Ling2007}; such solutions introduce distortions in addition to an increase in computational cost when the embedding dimension becomes larger than four \cite{Cuturi2013}. Additionally, they are not designed to be scalable to take advantages of large-scale GPU resources. \cite{Cuturi2013} has suggested improving the scalability of the distance calculation by using an iterative diagonal scaling algorithm, known as Sinkhorn's algorithm or iterative proportional fitting. We leave this scalability issue for future work. 

Similarly, computing the geodesic distance between two co-variance matrices is equally time inefficient -- $\mathcal{O}(4{d^3})$. The main component of this inefficiency emerges from the generalised eigenvalue equation, particularly for calculating multiple Cholesky factorisations each time a query is initiated. One way forward may be to use Stein's distance \cite{Sra2015} while preserving affine invariance and geometric properties inherited by the covariance matrices. Another way ahead is to perform the factorisation on a GPU \cite{Macindoe2013}. This becomes increasingly important if our framework were to be used for indexing of videos (instead of images). This future application relies on returning a set of similar videos in response to a query video. This could replace the current text based tagged video framework, like that used by several online video platforms, with feature based tagged videos.

\bibliography{cortexica_retrieval}
\bibliographystyle{ieee}

%\begin{appendices}
%\section{Appendix}
%\label{appendix:metric}
%
%
%\textbf{Definition of a metric:} If $\mathcal{D}$ characterizes the distance between two objects it is a metric \textit{iff} the following axioms hold: (a) $\mathcal{D} \left( {a,b} \right) \geqslant 0$, (b) $\mathcal{D}\left( {a,a} \right) = 0$, (c) $\mathcal{D}\left( {a,b} \right) = 0{\text{ if }}a = b$, (d) $\mathcal{D}\left( {a,b} \right) = \mathcal{D}\left( {b,a} \right)$ and (e) $\mathcal{D}\left( {a,c} \right) \leqslant \mathcal{D}\left( {a,b} \right) + \mathcal{D}\left( {b,c} \right)$. Semi-metrics only adhere to (a-b) and (d).
%\end{appendices}

\end{document}